\documentclass[10pt, a4paper]{article}

\usepackage{amsmath,amssymb}
\usepackage{hyperref}
\usepackage{bbm}
\usepackage{graphicx}
\usepackage{xcolor}
\usepackage{mathrsfs}
\usepackage{multicol}

\newcommand{\bw}{\begin{widetext}}
\newcommand{\ew}{\end{widetext}}

\newcommand{\be}{\begin{equation}}
\newcommand{\ee}{\end{equation}}
\newcommand{\bestar}{\begin{equation*}}
\newcommand{\eestar}{\end{equation*}}

\newcommand{\bi}{\begin{itemize}}
\newcommand{\ei}{\end{itemize}}
\newcommand{\bea}{\begin{eqnarray}}
\newcommand{\eea}{\end{eqnarray}}
\newcommand{\nin}{\noindent}

\newcommand{\hbo}{\hbox to 1 true cm {\hfill } }

\newcommand{\Prad}{P_{\mathrm{rad}}}
\newcommand{\sTh}{\sigma_{\mathrm{Th}}}
\newcommand{\sKN}{\sigma_{\mathrm{KN}}}

\newcommand{\sfrac}[2]{{\textstyle \frac{#1}{#2}}}

\newcommand{\vc}[1]{\mbox{\boldmath$#1$}}

\newcommand{\vcb}[1]{\mbox{\bf #1}}

\def\lambdabar{\protect\@lambdabar}
\def\@lambdabar{%
\relax \bgroup
\def\@tempa{\hbox{\raise.73\ht0
\hbox to0pt{\kern.2\wd0\vrule width.7\wd0
height.1pt depth.1pt\hss}\box0}}%
\mathchoice{\setbox0\hbox{$\displaystyle\lambda$}\@tempa}%
{\setbox0\hbox{$\textstyle\lambda$}\@tempa}%
{\setbox0\hbox{$\scriptstyle\lambda$}\@tempa}%
{\setbox0\hbox{$\scriptscriptstyle\lambda$}\@tempa}%
\egroup }

\begin{document}

\nin
{\Large \textbf{\sffamily Corrections to Laser Electron Thomson Scattering}}

\vspace{.5cm}

\nin

\vspace{.5cm}

\begin{multicols}{2}
\nin
\textbf{Thomas Heinzl} \\ School of Computing and Mathematics\\ Plymouth University \\
Drake Circus, Plymouth PL4 8AA, UK \\ theinzl@plymouth.ac.uk

\nin
\textbf{Anton Ilderton} \\ Department of Applied Physics\\ Chalmers University of Technology\\ SE-412 96 Gothenberg, Sweden \\ anton.ilderton@chalmers.se

\end{multicols}

We discuss classical and quantum corrections to Thomson scattering between an electron and a laser. For radiation reaction, nonlinear, and quantum effects we identify characteristic dimensionless parameters in terms of which we determine the leading order correction terms. 


\vspace{.5cm}

\begin{multicols}{2}

\section{Introduction}
Thomson scattering is the classical process of light being deflected by a charged obstacle of typical size smaller than the wavelength of  the incident radiation. For definiteness we will henceforth assume the charge to be an electron and the incoming wave stemming from a laser beam. The scattered radiation may be viewed as resulting from a two-step process: (i) the acceleration of the charge due to the electromagnetic field it encounters and (ii) the bremsstrahlung emitted in consequence. (In fact, this is what happens when solving the classical equations in powers of the coupling: at zeroth order, there is only acceleration, while at first order, there is emission.) The instantaneously radiated power is then given by Larmor's formula which, in its nonrelativistic incarnation and using Heaviside-Lorentz units, reads \cite{Schwinger}
\be \label{P.RAD.NR}
  \Prad = \frac{2}{3} \frac{e^2}{4\pi c^3} \dot{\vcb{v}}^2 \; .
\ee
Expressing the acceleration by means of the nonrelativistic Lorentz force (neglecting the $\vcb{v} \times \vcb{B}$ term), $m \dot{\vcb{v}} = e \vcb{E} $, $\vcb{E}$ being the electric field of the incoming  wave, the radiated power becomes
\be
  \Prad = \frac{2}{3} \frac{e^4}{4\pi m^2 c^3} E^2 \; .
\ee
This may be turned into a cross section (units of area) upon dividing by the energy flux, i.e.\ the modulus of the Poynting vector, which for a \emph{plane wave} (PW) is
\be \label{S.PW}
  S = c \, |\vcb{E} \times \vcb{B}| \stackrel{\mathrm{PW}}{=} c E^2 \; .
\ee
Hence, we obtain the Thomson cross section,
\be \label{THOMSON}
  \sTh := \frac{\Prad}{S} = \frac{8\pi}{3} \left(  \frac{e^2}{4\pi mc^2}\right)^2 =: \frac{8\pi}{3} r_e^2 \; ,
\ee
with the classical electron radius, $r_e \simeq 3$ fm. At this distance, the Coulomb energy between electrons equals $mc^2$, and (with the benefit of hindsight) one ends up with a distance scale that is actually typical for strong interactions (1 fm being roughly the nucleon size). As a result, the Thomson cross section takes on the numerical value $\sTh \simeq 6.6 \times 10^{-24}$ cm$^2 \simeq 0.66$ barn which is not small by particle physics standards.

The differential cross section may be derived in a similar vein. For this one needs the angular distribution of the radiated power (\ref{P.RAD.NR}) in direction $\vcb{l}$ which is \cite{Schwinger}
\be
  \frac{d\Prad}{d\Omega} = \frac{e^2}{(4\pi)^2 c^3} (\vcb{l} \cdot \dot{\vcb{v}})^2 \; ,
\ee
or, upon employing the equation of motion, with $\vcb{E} = E \vc{\epsilon}$,
\be
  \frac{d\Prad}{d\Omega} = \frac{e^4}{(4\pi)^2 m^2 c^3} E^2 \left( 1 - (\vcb{l} \cdot \vc{\epsilon})^2 \right) \; .
\ee
Averaging over polarisations and introducing the scattering angle $\theta$ one has
\be
  \langle 1 - (\vcb{l} \cdot \vc{\epsilon})^2 \rangle_{\mathrm{pol}} = \frac{1 + \cos^2 \theta}{2} \; .
\ee
This angular dependence was originally found by Thomson when calculating the mean ``rate at which energy is streaming through unit area'' \cite{Thomson:1903}.
Dividing by the incoming flux, $S = cE^2$, finally yields the differential cross section,
\be \label{DIFF.THOMSON}
  \frac{d\sTh}{d\Omega} = \sfrac{1}{2} r_e^2 \, (1 + \cos^2 \theta) \; .
\ee
Integrating over angles we reobtain (\ref{THOMSON}). Clearly, the differential cross section is more useful in general as it conveys more, namely spectral, information.

It is instructive to rederive (\ref{THOMSON}) in a fully covariant way. The radiated power may be inferred from the proper time derivative of the wave 4-momentum which can be expressed in terms of the electron 4-velocity $u^\mu$ \cite{Rohrlich:2007},
\be\label{P.RAD.REL0}
  \dot{P}^\mu = - \frac{2}{3} \frac{e^2}{4 \pi c^5} \dot{u}^2 u^\mu \; .
\ee
Dotting in $u$, using $u^2 = c^2$ and the covariant equation of motion, $m \dot{u}^\mu = (e/c)F^{\mu\nu} u_\nu$, one finds the radiated power,
\be \label{P.RAD.REL}
  \Prad = u \cdot \dot{P} = \sTh u_\mu F^{\mu\alpha} F_\alpha^{\; \nu} u_\nu /c =: \sTh w_0 \; .
\ee
For a plane wave (or, more general, null field), the square of the field strength tensor coincides with the energy-momentum tensor,
\be \label{T.MU.NU.PW}
  F^{\mu\alpha} F_\alpha^{\; \nu} \stackrel{\mathrm{PW}}{=} c \, T^{\mu\nu} \; ,
\ee
so that $w_0$ in (\ref{P.RAD.REL}) may be interpreted as the energy flux density of the electromagnetic wave as `seen' by the electron in its instantaneous rest frame.  This together with (\ref{P.RAD.REL}) yields a nice expression for the Thomson cross sections in terms of Lorentz invariants,
\be
  \sTh = \frac{u \cdot \dot{P}}{u_\mu T^{\mu\nu} u_\nu} \; ,
\ee
the numerator and denominator being the field energy rate of change and flux density in the co-moving frame of the electron, respectively.

\section{Radiation reaction}

The two-step procedure mentioned in the introduction breaks down (or rather, becomes insufficient) when the energy loss due to radiation becomes substantial such that the produced radiation field back-reacts on the particle motion. In this case, one has to solve a modified equation of motion taking back-reaction into account and named after Lorentz, Abraham and Dirac (LAD) \cite{Lorentz:1906, Abraham:1905, Dirac:1938}. Introducing the time parameter
\be
  \tau_0 := \frac{2}{3}r_e/c \simeq 10^{-23}  \text{s} \; ,
\ee
it is most compactly written as \cite{Rohrlich:2007} 
\be \label{LAD}
  m \dot{u}^\mu = F^\mu + \tau_0 \, m \ddot{u}^\mu =\frac{e}{c} F^{\mu\nu}u_\nu + \tau_0 \, m \ddot{u}^\mu \; .
\ee
This equation, being third order in time derivatives ($\ddot{u} = \dddot{x}$), has pathological features such as runaway solutions and preacceleration which can be traded for each other but not entirely removed while insisting on (\ref{LAD}) -- see the lucent discussion in \cite{Coleman:1961zz}. The loophole is to modify the equation via iteration, i.e.\ by expressing $m \ddot{u}$ on the right-hand side in terms of the Einstein-Lorentz force, $m \ddot{u} = \dot{F}^\mu + O(\tau_0)$ which yields the Landau-Lifshitz equation \cite{Landau:1987}
\be \label{LL}
  m \dot{u}^\mu =  F^\mu + \tau_0 \, m \dot{F}^\mu + O(\tau_0^2) \; .
\ee
This equation has been rederived (but only to order $\tau_0$) using adiabatic perturbation theory \cite{Spohn:1999uf} or a sophisticated classical regularisation procedure \cite{Gralla:2009md}. For plane wave backgrounds there is an exact analytic solution \cite{diPiazza:2008}

Clearly, either equation must lead to a modification of the Thomson cross section. This has already been worked out in Dirac's seminal paper \cite{Dirac:1938} who found the following simple result,
\be \label{SIGMA.RR1}
  \sigma_{\mathrm{RR}} = \frac{\sTh}{1 + (\omega_0 \tau_0)^2} \; .
\ee
Here we have introduced the laser frequency in the instantaneous rest frame, $\omega_0$. For a plane wave depending only on the invariant phase $k \cdot x$ with $k^2=0$, we can write $\omega_0 = k \cdot u$. In this case the solution of (\ref{LL}) implies $\omega_0 = k \cdot u = k \cdot u_0 + O(\tau_0)$ where $u_0$ is the \emph{initial} four-velocity \cite{diPiazza:2008}. Thus, $\omega_0$ is only conserved in the absence of radiation reaction \cite{Harvey:2011dp}. 

Assuming that our equations of motion both receive corrections of order $\tau_0^2$ we should rewrite the cross section as
\be \label{SIGMA.RR2}
  \sigma_{\mathrm{RR}} = \sTh \, \left( 1 - \omega_0^2 \tau_0^2 \right) \; .
\ee
The cross section (\ref{SIGMA.RR1}) may alternatively be obtained by considering an electron scattering off a bound charge, with the interaction described by an additional harmonic damping force, $\vcb{F}_{\mathrm{RR}} = - \delta m \vcb{v}$. The cross section is (see \cite{Schwinger}, Ch.\ 45),
\be
  \sigma_{\mathrm{RR}} = \frac{\sTh}{1 + \delta^2/\omega_0^2} \; ,
\ee
which we can match to (\ref{SIGMA.RR1}) by identifying
\be
  \delta = \omega_0^2 \tau_0 \; .
\ee
%
%
%
%
%
%
For a later comparison with the quantum effects it is useful to rewrite the small parameter $\omega_0 \tau_0$ in terms of the fine structure constant, $\alpha = e^2/4\pi \hbar c$, and the dimensionless parameter 
\be \label{NU0}
  \nu_0 := \frac{\hbar k \cdot u_0}{m c^2} = \frac{\hbar \omega_0}{mc^2} = \frac{\hbar e^\zeta \omega}{mc^2} \; ,
\ee
which measures the energy of the laser photons (as seen by the initial-state electron) in units of the electron rest mass. In the last expression we have introduced rapidity $\zeta$ via the electron gamma factor, $\exp{\zeta} = \gamma (1 + \beta)$, to write $\nu_0$ in terms of the lab frequency, $\omega$.  Obviously, one has
\be
  \omega_0 \tau_0 = \frac{2}{3} \alpha \nu_0 \; ,
\ee
where the factors of $\hbar$ cancel on the right-hand side as they must for a purely classical parameter. Let us estimate the magnitude of this quantity for a relativistic electron ($\beta \simeq 1$) colliding with an optical laser \cite{Harvey:2011dp,Harvey:2010ns}. In this case one has $\nu := \hbar \omega/mc^2 \simeq 10^{-6}$ and $e^\zeta \simeq 2 \gamma$. Thus, $\omega_0 \tau_0 \simeq \alpha \gamma \nu \simeq 10^{-8} \gamma$, which implies that classical radiation reaction will be substantial for electron energies of $10^2$ TeV which is in the far quantum regime. So, unless we find a way of \emph{classically} boosting the radiation reaction its observation will not become feasible. A possible means towards this end is nonlinearity \cite{koga:2005,DiPiazza:2009zz}.

\section{Nonlinear Effects}

Laser intensity is measured in terms of the r.m.s. energy gained by an electron traversing a laser wave length, in units of the electron rest energy. This leads to the dimensionless laser amplitude
\be
  a_0 = \frac{eE_{\mathrm{rms}} \lambdabar}{mc^2} = \frac{eE_{\mathrm{rms}}}{mc\omega}\; ,
\ee
which may be expressed covariantly using the energy density of (\ref{P.RAD.REL}) such that \cite{Heinzl:2008rh}
\be
  a_0^2 = \frac{e^2 \langle u_\mu T^{\mu\nu} u_\nu \rangle_{\mathrm{rms}}}{m^2 c^2 (k \cdot u)^2} \; .
\ee
When the laser beam is focussed down to the diffraction limit this may be turned into a rule-of-thumb estimate of $a_0$ in terms of laser power $P_L$ measured in petawatts (PW), namely
\be
  a_0^2 \simeq 5 \times 10^{3} \, P_L/\mbox{PW} \; .
\ee
For powers of 10 PW to be expected in the near future this implies $a_0$ values of a few hundred. For the sake of simplicity, however, we will restrict our interest to the nonlinear corrections of leading order in $a_0$.

It is known that an electron in a circularly polarised plane wave moves along a circle (in the average rest frame where there is no longitudinal drift, see \cite{Landau:1987}, Ch.\ 48, Problem 3). The radiation spectrum for such a charge motion has been first calculated by Schott \cite{Schott:1912} [formula (128), Sect.\ VII.84; see also \cite{Schwinger}, Ch.\ 38] and consists of an infinite sum of higher harmonic contributions labelled by an integer $n$. Turning the $n$th harmonic intensity scattered per solid angle into a differential cross section yields \cite{McDonald:1986zz}
\be \label{SCHOTT}
  \frac{d\sigma_{n}}{d\cos \theta} = \frac{4\pi n r_e^2}{a_0^2} \left( \cot^2 \theta \, J_n^2 (n a_0 \sin \theta) + a_0^2 J_n^{\prime 2} (n a_0 \sin \theta) \right)
\ee
where $n$ is the harmonic number and $\theta$ the scattering angle. To find the leading correction to linear Thomson scattering ($n=1$, $a_0 = 0$) we expand the Bessel functions for $a_0 \ll 1$ using the first terms in their power series,
\be
  J_n (z) = (z/2)^n \left\{ \frac{1}{n!} - \frac{z^2/4}{(n+1)!} + O (z^4) \right\}  \; .
\ee
Plugging this into (\ref{SCHOTT}) for $n=1$ results in the differential cross section
\be \label{D.SIGMA1}
  \frac{d\sigma_{1}}{d\cos \theta} = r_e^2 \pi \left( 1 + \cos^2 \theta  - \sfrac{1}{4} a_0^2 \sin^2 \theta \, (3 + \cos^2 \theta) \right) \; ,
\ee
which nicely exhibits the correction to the Thomson cross section (\ref{DIFF.THOMSON}) of order $a_0^2$. Integrating over the angle $\theta$ we find the total cross section (for the fundamental harmonic, $n=1$),
\be \label{SIGMA1}
  \sigma_1 = \sTh \left( 1 - \sfrac{2}{5} a_0^2 \right) \; .
\ee
As $J_n^2 (n a_0 \sin \theta) \sim a_0^{2n}$ there is another $a_0^2$ correction coming from the second harmonic. Expanding for $n=2$ one obtains
\be \label{D.SIGMA2}
  \frac{d\sigma_{2}}{d\cos \theta} = 2 \, r_e^2 \pi \, a_0^2 \sin^2 \theta \left( 1 +  \cos^2 \theta \right) \; .
\ee
Interestingly, this has the same overall $\sin^2 \theta$ angular dependence as the $a_0^2$ correction in the fundamental cross section (\ref{D.SIGMA1}), and hence the two will be difficult to distinguish experimentally. This statement holds for circular polarisation, but for \emph{linear} polarisation the situation is different and higher harmonics have been unambiguously identified from their angular radiation patterns \cite{Chen:1998}.

\includegraphics[width=0.8\columnwidth]{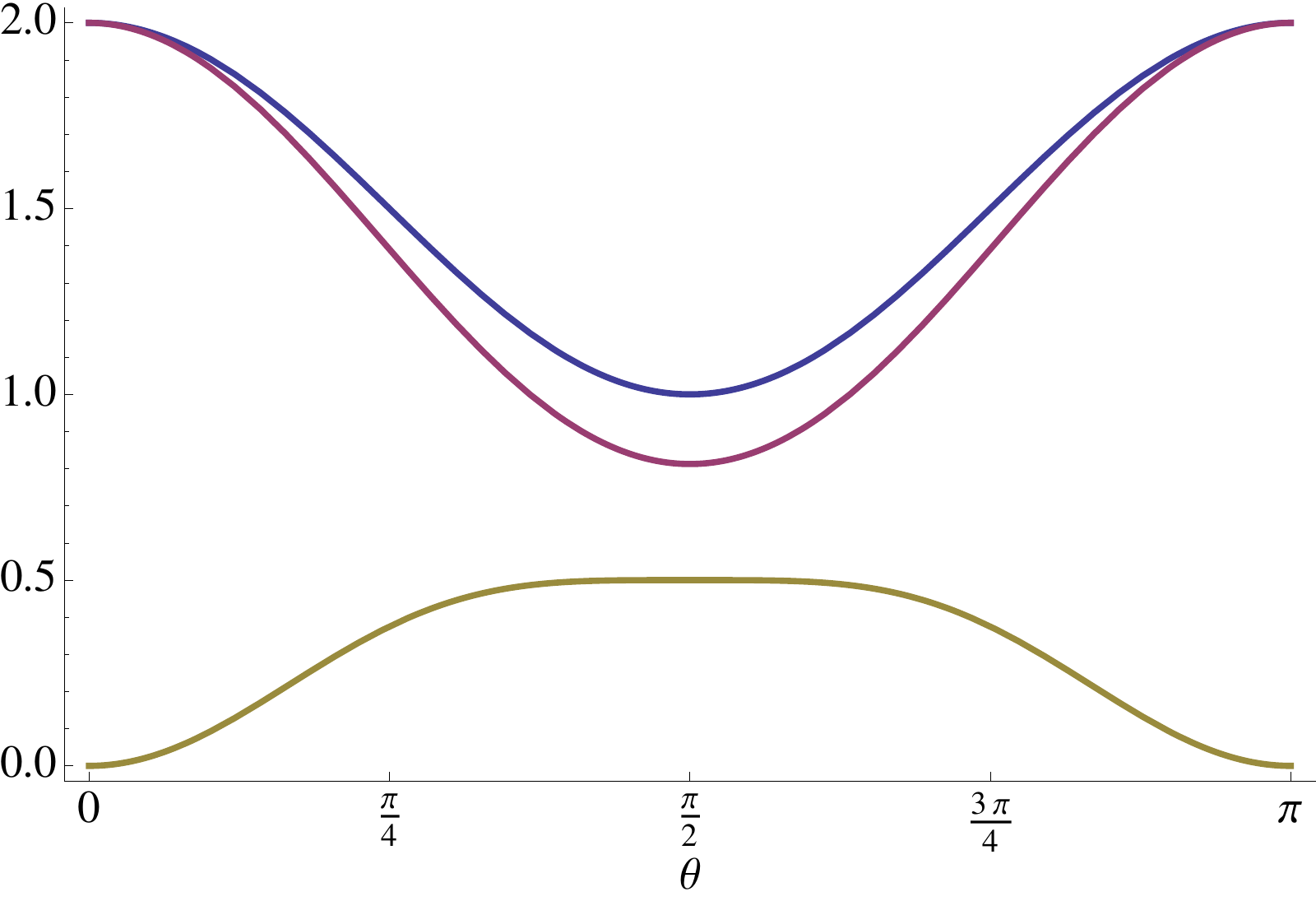}

\nin
Fig.~1: $d\sTh/d\theta$ (upper curve), $d\sigma_1/d\theta$ (middle curve) and $d\sigma_2/d\theta$ (lower curve), in units of $r_e^2 \pi$, for $a_0 = 0.5$, cf.\ (\ref{DIFF.THOMSON}), (\ref{D.SIGMA1}) and (\ref{D.SIGMA2}).

\vspace{.5cm}

\nin
The angular dependence on $\theta$ of both cross sections, (\ref{D.SIGMA1}) and (\ref{D.SIGMA2}) are compared with the Thomson one (\ref{DIFF.THOMSON}) in Fig.~1. Note that only the Thomson term ($a_0=0$) contributes in forward and backward direction ($\theta=0$ and $\theta=\pi$, respectively) as the $a_0^2$ corrections are suppressed there by the $\sin^2 \theta$ factor.

Integrating (\ref{D.SIGMA2}) over $\theta$ results in a contribution of different sign
\be \label{SIGMA2}
  \sigma_2 = \sfrac{6}{5} \, a_0^2 \, \sTh \; .
\ee
As a consequence,  upon adding (\ref{SIGMA1}) and (\ref{SIGMA2}), the complete order $a_0^2$ correction in the total cross section, i.e.\ in the sum
\be
  \sigma = \sigma_1 + \sigma_2 = \sTh \left( 1 + \sfrac{4}{5} a_0^2 \right) \; ,
\ee
becomes positive.

\section{Quantum Effects}

The quantum version of Thomson scattering is of course Compton scattering. Quantum corrections become important when the energy of the radiation becomes comparable to the electron rest mass. In this case, there will be substantial transfer of four-momentum and the whole scattering process is most naturally described using the photon picture, that is, by considering the process $e(p) + \gamma(k) \to e'(p') + \gamma'(k')$.  Energy momentum conservation then reads $k+p = k' + p'$ and, in the technically simplest case with the electron initially at rest, implies the Compton formula for the scattered frequency,
\be \label{OOPRIME}
  \frac{\omega}{\omega'} = 1 +  \nu_0 (1 - \cos \theta) \; ,
\ee
with the invariant $\nu_0$ defined in (\ref{NU0}). In the classical limit, $\nu_0 \to 0$, the energy transfer (recoil) vanishes and $\omega = \omega'$. The differential cross section has first been obtained by Klein and Nishina \cite{KleinNishina:1929} and reads \cite{Berestetskii:1982}
\be \label{KLEIN.NISHINA}
  \frac{d\sKN}{d\Omega} = \sfrac{1}{2} r_e^2 \left( \frac{\omega'}{\omega} \right)^2 \left( \frac{\omega}{\omega'} + \frac{\omega'}{\omega} - \sin^2 \theta \right) \; .
\ee
The classical Thomson limit (\ref{DIFF.THOMSON}) is readily obtained by setting $\omega = \omega'$. Integrating (\ref{KLEIN.NISHINA}) over angles and expanding for $\nu_0 \ll 1$ we obtain the leading quantum correction to the Thomson  cross section (\ref{THOMSON}),
\be \label{SKN}
  \sKN \simeq \sTh (1 - 2 \nu_0) \; .
\ee

\section{Synthesis}

Ideally, one would like to have a general theory that incorporates all corrections. In principle, such a theory exists (at least to some extent), namely quantum electrodynamics (QED) coupled to a strong external plane wave field, a particular incarnation of strong-field QED. It was originally developed in a series of papers, the most relevant ones being \cite{Toll:1952,Reiss:1962,Nikishov:1963,Brown:1964zzb}. The main challenge for the present discussion is to go beyond the plane wave model and describe the external laser field, say by a Gaussian beam. In this case one loses the exact solution of the Dirac equation due to Volkov \cite{Volkov:1935}, and basically no progress has been made along this route. So one will have to make do with plane wave backgrounds for the time being.

Nevertheless, important achievements have been reported. For the case of circular polarisation, the most comprehensive collection of results can be found in \cite{Narozhnyi:1964} (see also Ch.\ 101 of \cite{Berestetskii:1982}), in particular the strong field QED generalisation of the classical result (\ref{SIGMA1}) which usually is referred to as `nonlinear Compton scattering' (NLC). We will not repeat the lengthy formulae here, but only present our result for the expansion of the fundamental cross section $\sigma_1$ which reads,
\be \label{NLC1}
  \sigma_{\mathrm{NLC},1} = \sTh \left( 1 - 2 \nu_0 - \sfrac{2}{5} a_0^2 + \sfrac{14}{5} \nu_0 a_0^2 + \ldots \right) \; .
\ee
Comparing with (\ref{SIGMA1}) we find the same nonlinear correction of order $a_0^2$ \emph{plus} the quantum correction $- 2 \nu_0$ from (\ref{SKN}). This provides a useful consistency check. We also note that higher order corrections mix nonlinear and quantum contributions such as the last term of order $\nu_0 a_0^2$ in (\ref{NLC1}). In this rather literal sense, the NLC result (\ref{NLC1}) may be viewed as a unification of quantum and nonlinear corrections to Thomson scattering. What seems to be missing in this unification are the radiation reaction contributions. To incorporate these, one must understand how the classical RR corrections (\ref{SIGMA.RR1}) and (\ref{SIGMA.RR2}) arise, and then how they can be derived from QED.

The classical result (\ref{SIGMA.RR1}) comes from inserting the orbit of a radiating electron into (\ref{P.RAD.REL0}) and (\ref{P.RAD.REL}) to calculate the emitted power. That orbit is calculated, classically, by eliminating the gauge field. The analogous quantum calculation would involve summing over all possible numbers of outgoing photons  \cite{DiPiazza:2010mv,DiPiazza:2011tq,DiPiazza:2013iwa}. The NLC cross section (\ref{NLC1}) contains all quantum and intensity-dependent corrections to Thomson scattering, but only in the process of {\it one}-photon emission (from the collision of an electron and a laser). To obtain the full result (\ref{SIGMA.RR1}) one would need to calculate an appropriate quantity in QED to all orders, which is challenging. Low order corrections are more readily calculated; radiation reaction appears in the scattered electron momentum following single photon emission, and in the radiated photon momentum at higher orders, following e.g.\ two-photon emission~\cite{Ilderton:2013tb,Ilderton:2013dba}. The lowest order term of Dirac's result should therefore come from the two-photon emission diagram.

All this suggests that the full theory should contain mixed terms of the form $(\alpha \nu_0 a_0)^{2n}$ implying the possibility of a regime ($a_0 \gg 1$) where quantum effects are still moderate but radiation reaction gets boosted due to large nonlinearities.

%
%

\bibliographystyle{plain}

\end{multicols}

\end{document}